\documentclass[preprint,pre,a4paper,superscriptaddress,floatfix,longbibliography]{revtex4-1}
\usepackage{graphicx,color,epsfig,subfig,dcolumn,bm,mathrsfs,amsmath,amssymb}
\usepackage[colorlinks]{hyperref}
\setcitestyle{super}

\newcommand{\change}[1]{\textcolor{black}{#1}}

\begin{document}

\title{Directional freezing of binary colloidal suspensions: A model for size fractionation of graphene oxide}
\author{Xin Xu}
\affiliation{School of Chemistry, Key Laboratory of Bio-Inspired Smart Interfacial Science and Technology of Ministry of Education, Beihang University, Beijing 100191, China}
\affiliation{Center of Soft Matter Physics and its Applications, Beihang University, Beijing 100191, China}
\author{Luofu Liu}
\affiliation{School of Chemistry, Key Laboratory of Bio-Inspired Smart Interfacial Science and Technology of Ministry of Education, Beihang University, Beijing 100191, China}
\affiliation{Center of Soft Matter Physics and its Applications, Beihang University, Beijing 100191, China}
\author{Hongya Geng}
\affiliation{Key Laboratory of Green Printing, Institute of Chemistry, Chinese Academy of Sciences, Beijing 100190, China}
\author{Jianjun Wang}
\affiliation{Key Laboratory of Green Printing, Institute of Chemistry, Chinese Academy of Sciences, Beijing 100190, China}
\author{Jiajia Zhou}
\email[]{jjzhou@buaa.edu.cn}
\affiliation{School of Chemistry, Key Laboratory of Bio-Inspired Smart Interfacial Science and Technology of Ministry of Education, Beihang University, Beijing 100191, China}
\affiliation{Center of Soft Matter Physics and its Applications, Beihang University, Beijing 100191, China}
\affiliation{Beijing Advanced Innovation Center for Biomedical Engineering, Beihang University, Beijing 100191, China}
\author{Ying Jiang}
\email[]{yjiang@buaa.edu.cn}
\affiliation{School of Chemistry, Key Laboratory of Bio-Inspired Smart Interfacial Science and Technology of Ministry of Education, Beihang University, Beijing 100191, China}
\affiliation{Center of Soft Matter Physics and its Applications, Beihang University, Beijing 100191, China}
\affiliation{Beijing Advanced Innovation Center for Biomedical Engineering, Beihang University, Beijing 100191, China}
\author{Masao Doi}
\affiliation{Center of Soft Matter Physics and its Applications, Beihang University, Beijing 100191, China}

\begin{abstract}
The performance of graphene oxide(GO)-based materials strongly depends on the
lateral size and size distribution of GO nanosheets. 
Various methods are employed to prepare GO nanosheets with a narrow size distribution. 
One of the promising method was proposed recently by directionally freezing of a GO aqueous dispersion at a controlled growth rate of the freezing front. 
We develop a theoretic model of binary colloids suspension, incorporating both the moving freezing boundary and the preferential adsorption of colloidal particles to the ice phase. 
We numerically solve the coupled diffusion equations and present state diagrams of size fractionation. 
Selective trapping of colloids according to their size can be achieved by a suitable choice of the experimental parameters, such as the adsorption rates and the freezing speed. 
\end{abstract}

\maketitle

\section{Introduction}

The size of graphene oxide (GO) nanosheets plays an extraordinarily crucial role in controlling the microstructures and properties of GO-based materials. 
This size tunability of the material properties offers great potential for applications in material science and nanotechnology. 
The distinct sizes of GO can demonstrate their individual merits. 
Large GO sheets are desirable for fabricating conductive thin films for optoelectronic devices.\cite{ChangHaixin2010, ZhengQingbin2011} 
On the contrary, small GO sheets are favorable for sensing and biological applications.\cite{LiuZhuang2008, Agarwal2010} 
Furthermore, the sizes of GO sheets play an important role in controlling the electrical and thermal conductivities of GO-based materials.\cite{Nika2009, Ghosh2010}

In practice, however, solutions of GO nanosheets usually have a broad size distribution due to particular preparation processes. 
Therefore, a tremendous variety of chemical and physical size fractionation methods have been proposed and developed to achieve the GO sheets with the relatively narrow distribution in size. 
Shi et al. designed a size separation method based on the pH-dependent amphiphilicity of GO sheets via adjusting the pH value of dispersion.\cite{WangXiluan2011} 
Taking into account of technical difficulties of precisely adjusting the pH values of GO solutions, they further developed a method based on the exclusive diffusion of GO sheets from its aqueous dispersion through precisely size-defined pores of track-etched membranes.\cite{ChenJi2015} 
By virtue of the peculiar dispersibility and stability of GO sheets in polar solvent, a polar solvent-selective sedimentation method was developed by Zhang et al. to obtain series of products with size homogeneity.\cite{ZhangWenjun2014} 
Sun et al. conducted a density gradient ultracentrifugation rate separation method to sort GO sheets by taking advantage of their differences in sedimentation rate.\cite{SunXiaoming2010} 
Later, Lee et al. developed a facile size selection method based on the spontaneous accumulation of large-size GO flakes within the liquid crystalline phase.\cite{Lee2014} 
Nevertheless, these methods mentioned above have their individual drawbacks in practice, such as requirement of special equipment, sophisticated operating procedures to remove the chemical additives, time or energy consuming cost.

Recently, we developed a simple and efficient method to achieve the size fractionation of GO nanosheets by directionally freezing a GO aqueous dispersion at a controlled growth rate of the freezing front. \cite{GengHongya2017}
The stratification of GO sheets in the solution leads to a preferential adsorption of small GO sheets in the ice phase, and the final results show a narrow size distribution of GO sheets with the size ranging from several nanometers to tens of micrometers. 
This typical non-equilibrium phenomenon is simultaneously controlled by several factors, such as the size distribution of GO sheets, the freezing rate of ice front, and the preferential adsorption of GO sheets to the ice phase. 
However, the underlying mechanism for the size fractionation of GO sheets is still not clear.

Here, we shall provide a detailed numerical study of the directional freezing of binary colloidal suspensions. 
We make a crude simplification and model the GO sheet as a hard sphere of similar size.
The spherical approximation should be valid for nanometer-size GO sheets due to the rapid rotational motion in comparison to their \change{translational} Brownian motion, but is questionable for GO sheets of micrometer size.
Here we proceed with this approximation and show later that a qualitative agreement with the experiments can be achieved. 
In the present work, we primarily focus on the freezing front moving with a low velocity when the instability of the freezing front does not happen.\cite{Peppin2008} 
In this case, the ice front can be regarded as a planar-like interface, and the evolution of the system be reduced to be a one-dimensional problem. 
Furthermore, the slow freezing rate allows us to consider the trapping of colloids in the ice phase as a quasi-equilibrium process, i.e., the preferential trapping of colloids is governed by the equilibrium of chemical potentials between ice phase and liquid phase. 
More complete treatment involves the intricate competition from many types of forces.\change{\cite{Rempel1999, Rempel2001, Saint-Michel2017, YouJiaxue2018}} 
In addition, the dilute limit we concern here ensures the validity of the two-body interaction approximation describing the cross-section interaction which has been demonstrated as the crucial factor in controlling the stratification of binary colloids in drying films.\cite{ZhouJiajia2017, ZhouJiajia2017a}

The remainder of this article is organized as follows: 
In Section \ref{sec:model}, we derive the dynamical equations that describe the evolution of binary colloids with a moving freezing front. 
We then numerically solve these equations, and plot state diagrams of size fractionation by varying the system parameters, such as the size ratio, the freezing rate, and the difference in the chemical potentials of colloids in ice/solution phases (Section \ref{sec:result}). 
We then undertake the comparison of our predictions with experimental observations and demonstrate the mechanism of fractionation.
Finally, we conclude in Section \ref{sec:summary} with a brief summary.
\\

\section{Theoretical Model}
\label{sec:model}

In this section, we set up a diffusion model that describes the distribution of colloidal particles in the ice and in the solution, as the freezing front moves from top to bottom. 
We focus on the case of slow freezing rates, where the advancing ice front keeps a planar profile\cite{Peppin2006, Peppin2008} and some colloids will be trapped in the ice due to adsorption. 
However, at faster freezing rates, a non-planar front arises\cite{Peppin2007, Deville2009} due to interfacial instability during the solidification of colloidal suspensions, particularly for the constitutional supercooling where the temperature of freezing front is below its equilibrium freezing temperature. 
The system out of equilibrium with non-planar interface is extremely complicated, related to the intricate interaction between the freezing front and colloids, and is beyond the scope of our current study. 
Here, we only consider the fractionation of the planar freezing front for colloidal mixtures. 
In addition, the lateral flow can also be ignored because of the relatively low concentration \change{we consider here}. 
Thus, our system can be regarded to freeze one dimensionally.
\change{The effect of the gravity is also neglected in our model.}

\begin{figure}
  \centering
  \includegraphics[scale=0.6,draft=false]{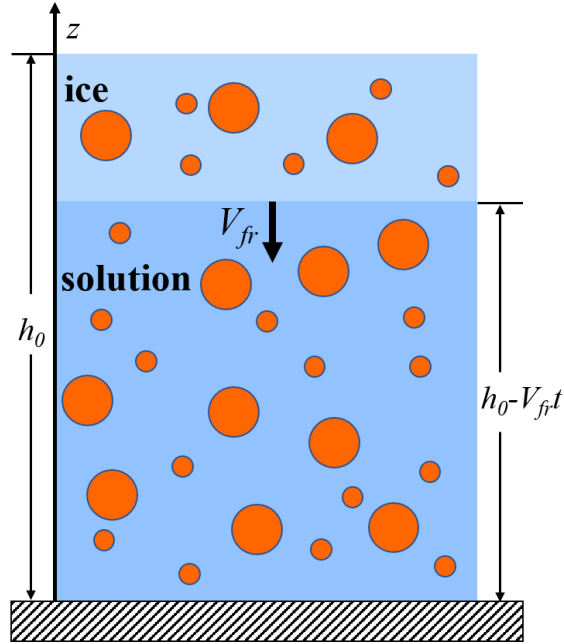}
  \caption{A colloidal solution is frozen from top to bottom, consisting of two types of particles with different sizes. \change{The initial height of the solution without ice is $h_0$}. As the freezing front moves along the $-z$ direction with a constant velocity $V_{fr}$, some particles are trapped in the ice and the height of residual solution is given by $h_0-V_{fr}t$.}
  \label{fig:1}
\end{figure}

We consider a colloidal solution that is composed of two types of particles with different sizes, as schematically shown in Fig. \ref{fig:1}. 
The solution is frozen from the top surface, i.e., $z=h_0$, and the freezing front moves downward along the $-z$ direction with a constant speed $V_{fr}$. 
The evolution of the solution height satisfies $h(t)=h_0-V_{fr}t$. 
The colloids are considered as hard spheres with the radius $r_S$ (small) and $r_L$ (large), and their volumes are $v_i=4\pi r_i^3/3 (i=S, L)$. 
We define the size ratio of colloids by $\alpha=r_L/r_S$. 
Because of the interaction between the freezing front and the colloids, some particles will be trapped inside the ice as the freezing front moves. 
The time-dependent volume fraction can be denoted by $\phi (t,z\in[0,h])$ in the solution and $\phi_{Ci} (t,z\in[h,h_0])$ in the ice. 
The number density can be related to volume fraction by $\phi_i=n_iv_i$. 
Initially, the volume fractions for both species in the solution are assumed to be homogeneous, i.e. $\phi_i (t=0,z)=\phi_{0i}$.

We shall derive the \change{evolution equations for the} colloidal \change{volume fractions} using the Onsager variational principle. 
This principle essentially states that the time evolution of the system is determined by the balance of two forces: the \change{thermodynamic} force that drives the system to the state of \change{equilibrium}, and the \change{viscous} frictional force that resists this change.\cite{DoiSoft} 
The evolution equations can be determined by the minimization of Rayleighian $\mathscr{R}=\dot{A}+\Phi$, where $\dot{A}$ is the rate of change of free energy $A$, and $\Phi$ is the dissipation function accounting for the frictional force exerted by the fluid when particles move.
  
The free energy can be expressed by
\begin{equation}
  \label{equ1}
  A=\int_0^h\mathrm dz \, f(\phi_S,\phi_L),
\end{equation}
where the free energy density for the hard sphere mixture reads
\begin{equation}
  \label{equ2}
  \dfrac{1}{k_BT}f(\phi_S,\phi_L) =\sum_{i=S,L}\dfrac{\phi_i}{v_i}\ln\phi_i+\sum_{i,j=S,L}\dfrac{a_{ij}}{v_iv_j}\phi_i\phi_j.
\end{equation} 
In the free energy expression (\ref{equ2}), we have \change{used the local density approximation (so no gradient terms come into the free energy) and kept up to the second-order virial expansion.\cite{HansenMcDonald}
These approximations should be valid for dilute solutions.}
In Eq. (\ref{equ2}), $k_B$ and $T$ denote the Boltzmann constant and absolute temperature, respectively. 
In addition, the second-order virial coefficient is given by $a_{ij}=({2\pi}/{3})(r_i+r_j )^3$. 

The number of particles is conserved, leading to the conservation equation
\begin{equation}
  \label{equ4}
  \dot\phi_i=\frac{\partial \phi_i}{\partial t}
  =-\frac{\partial (\phi_i\upsilon_i)}{\partial z},
\end{equation}
where $\upsilon_i$ is the averaged particle velocity.
The time derivative of free energy can be calculated using the integral by parts
\begin{equation}
  \change{\dot{A}} = \int_0^h\mathrm dz \sum_{i=S,L}\dfrac{\partial f(\phi_S,\phi_L)}{\partial \phi_i}\dot\phi_i 
  = -\sum_{i=S,L}\dfrac{\partial f(\phi_S,\phi_L)}{\partial \phi_i}\phi_i\upsilon_i\bigg|_0^h+\int_0^h\mathrm dz \sum_{i=S,L}\dfrac{\phi_i\upsilon_i}{v_i}\dfrac{\partial\mu_i}{\partial z} ,
\label{equ3}
\end{equation}
where the chemical potential $\mu_i$ is given by
\begin{equation}
  \label{equ5}
  \mu_i=\dfrac{\partial f(\phi_S,\phi_L)}{\partial n_i}
  = v_i \dfrac{\partial f(\phi_S,\phi_L)}{\partial \phi_i} . 
\end{equation}

The dissipation \change{function} is a quadratic function of $\upsilon_i$,
\begin{equation}
  \label{equ6}
  \Phi=\dfrac{1}{2} \int_0^h \mathrm dz 
  \sum_{i,j=S,L} \xi_{ij}\upsilon_i\upsilon_j ,
\end{equation}
where $\xi_{ij}$ is the friction matrix. 
In the present work, we ignore the hydrodynamic interaction and assume the friction matrix is diagonal: $\xi_{SL}=\xi_{LS}=0, \xi_{SS}=n_S \zeta_S$ and  $\xi_{LL}=n_L \zeta_L$, where $\zeta_i$  denotes the friction constant per particle.

Minimization of the Rayleighian $\mathscr{R}=\dot {A}+\Phi$ with respect to $\upsilon_i$ leads to the expression for the average velocities $\upsilon_i$
\begin{eqnarray}
  \label{equ9}
  \upsilon_S &=& -D_S\bigg[(\frac{1}{\phi_S}+8)\frac{\partial\phi_S}{\partial z}+(1+\frac{1}{\alpha})^3 \frac{\partial\phi_L}{\partial z}\bigg] , \\
  \label{equ10}
  \upsilon_L &=& -D_L\bigg[(\frac{1}{\phi_L}+8)\frac{\partial\phi_L}{\partial z}+(1+\alpha)^3 \frac{\partial\phi_S}{\partial z}\bigg] .
\end{eqnarray}
Here, the Einstein relation $D_i=k_B T/\zeta_i$ has been used. 

The evolution equations for the colloidal concentrations can be obtained by the substitution of Eq.~(\ref{equ4})
\begin{eqnarray}
  \label{equ7}
  \frac{\partial\phi_S}{\partial t} &=& D_S\frac{\partial}{\partial z}\bigg[(1+8\phi_S)\frac{\partial\phi_S}{\partial z}+(1+\frac{1}{\alpha})^3 \phi_S\frac{\partial\phi_L}{\partial z}\bigg] , \\
  \label{equ8}
\frac{\partial\phi_L}{\partial t} &=&D_L\frac{\partial}{\partial z}\bigg[(1+8\phi_L)\frac{\partial\phi_L}{\partial z}+(1+\alpha)^3 \phi_L\frac{\partial\phi_S}{\partial z}\bigg] .
\end{eqnarray}
Equations (\ref{equ7}) and (\ref{equ8}) are coupled diffusion equations which should satisfy a set of boundary conditions. 
At the bottom, i.e. $z=0$, the boundary condition is $\upsilon_i=0$. 
However, at the top surface of the solution i.e. $z=h$, the situations ranging from trapping and rejection are complicated due to the intricate interactions between dispersed colloids and the freezing front, such as thermodynamic forces and hydrodynamic forces.\change{\cite{Rempel1999, Rempel2001, YouJiaxue2018}} 
Nevertheless, regardless of the elaborate competition among these range-dependent forces, here, we take the simplest assumption that the quasi-static equilibrium condition is satisfied as the ice front moves at a slow speed. 
Then, we can determine this boundary condition by the equivalence of chemical potentials at interface between the ice phase and the suspension phase for the identical species
\begin{equation}
  \label{equ11}
  \mu_{Ci}(z \rightarrow h^+) = \mu_i(z \rightarrow h^-) .
\end{equation}
The chemical potential in the ice is
\begin{equation}
  \label{equ12}
  \mu_{Ci}(z \rightarrow h^+) = \mu_{Ci}^0+k_BT\bigg( \ln\phi_{Ci} +1\bigg) ,
\end{equation}
and its counterpart in the solution is
\begin{equation}
  \label{equ13}
  \mu_i(z \rightarrow h^-) = k_BT 
  \left( \ln\phi_i +1+2\sum_j \frac{a_{ij}}{v_j}\phi_j\right) .
\end{equation}
\change{Note here $\mu_{Ci}^0$ represents the change of the chemical potential for one colloids previous in the solution phase enters in the ice phase. 
Therefore, this quantity is related to the preference of colloids to either the ice or the solution.}

Substituting Eqs.~(\ref{equ12}) and (\ref{equ13}) into Eq.~(\ref{equ11}), we can obtain the boundary condition at the freezing interface $(z=h)$
\begin{equation}
  \label{equ14}
  \change{\phi_{Ci}(h)}=k_i\phi_i(h)\exp\left( 2\sum_j \frac{a_{ij}}{v_j}\phi_j(h) \right) ,\quad  k_i=\exp\left( -\frac{\mu_{Ci}^0}{k_BT}\right) .
\end{equation}
\change{Here $k_i$ is the Boltzmann factor of the quantity $\mu_{Ci}^0$. 
In the very dilute limit if one neglect the second-order virial contribution, $k_i=\phi_{Ci}(h) / \phi_i(h)$ gives the ratio of the colloidal concentrations in the ice phase and in the solution phase at the freezing front. 
When $\mu_{Ci}^0>0$, $k_i<1$ and $\phi_{Ci}(h)<\phi_{i}(h)$, the colloids prefer staying in the solution phase.
When $\mu_{Ci}^0<0$, $k_i>1$ and $\phi_{Ci}(h)>\phi_{i}(h)$, the colloids are adsorbed into the ice phase.}

Associating with \change{the condition that particle numbers are conserved across the interface} $\phi_i \upsilon_i=(\phi_{Ci}-\phi_i ) V_{fr}$, the boundary condition at the freezing interface reads
\begin{equation}
  \label{equ15}
  \upsilon_i \big|_{z=h} = V_{fr} \bigg[ k_i\exp \left( 
    2\sum_{j=S,L} \frac{a_{ij}}{v_j}\phi_j(h) \right) -1\bigg]
\end{equation}
Here the boundary conditions at the freezing front are coupled as well.

The coupled diffusion Eqs. (\ref{equ7}) and (\ref{equ8}) can be made dimensionless by scaling the length to the initial height of the solution $h_0$  and the time to the freezing time scale $\tau=t/(h_0/V_{fr})$. 
The dimensionless diffusion equations are
 \begin{eqnarray}
   \label{equ16}
     \frac{\partial\phi_S}{\partial \tau} + 
     \frac{x}{1-\tau} \frac{\partial\phi_S}{\partial x}
     &=& \frac{1}{{\rm Pe}_S(1-\tau)^2}
      \frac{\partial}{\partial x} 
     \bigg[(1+8\phi_S)\frac{\partial\phi_S}{\partial x} 
     +(1+\frac{1}{\alpha})^3 \phi_S\frac{\partial\phi_L}{\partial x}\bigg] , \\
  \label{equ17}
    \frac{\partial\phi_L}{\partial \tau} + 
    \frac{x}{1-\tau}\frac{\partial\phi_L}{\partial x}
    &=& \frac{1}{{\rm Pe}_L(1-\tau)^2} 
     \frac{\partial}{\partial x}
    \bigg[(1+8\phi_L)\frac{\partial\phi_L}{\partial x}
    +(1+\alpha)^3 \phi_L\frac{\partial\phi_S}{\partial x}\bigg] .
\end{eqnarray}

This procedure leads to two Peclet numbers
\begin{equation}\label{equ18}
{\rm Pe}_S=\frac{V_{fr} h_0}{D_S}, \quad {\rm Pe}_L=\frac{V_{fr} h_0}{D_L}=\alpha {\rm Pe}_S ,
\end{equation}
which quantify the competition between the freezing of the solution and the diffusion of the colloids. 
For $\mathrm{Pe}<1$, the system is diffusion-dominant and tends to exhibit a uniform distribution. 
On the contrary, for $\mathrm{Pe}>1$, the colloidal accumulation will lead to the creation of a concentration gradient.

In solving the coupled Eqs. (\ref{equ7}) and (\ref{equ8}), normal numerical routines are unstable when the concentrations occasionally become less than zero (also nonphysical). 
In order to avoid this \change{problem}, we perform the following substitution of variables
\begin{equation}\label{equ19}
\phi_S=e^{u_S},\quad \phi_L=e^{u_L}
\end{equation}
In this way, the positiveness of $\phi_S$ and $\phi_L$ are guaranteed. 
Then the dimensionless diffusion equations become
\begin{eqnarray}
  \label{equ20}
    e^{u_S} \frac{\partial u_S}{\partial \tau} 
    + e^{u_S} \frac{x}{1-\tau} \frac{\partial u_S}{\partial x} 
    & = & \frac{1}{{\rm Pe}_S(1-\tau)^2} \frac{\partial}{\partial x}
    \bigg[ (1+8e^{u_S}) e^{u_S} \frac{\partial u_S}{\partial x}
    + (1 + \frac{1}{\alpha})^3 e^{u_S} e^{u_L} \frac{\partial u_L}{\partial x}\bigg] , \\
  \label{equ21}
    e^{u_L} \frac{\partial u_L}{\partial \tau}
    + e^{u_L} \frac{x}{1-\tau} \frac{\partial u_L}{\partial x} 
    & = & \frac{1}{{\rm Pe}_L(1-\tau)^2} \frac{\partial}{\partial x}
    \bigg[ (1+8e^{u_L}) e^{u_L} \frac{\partial u_L}{\partial x}
    + (1+\alpha)^3 e^{u_S} e^{u_L} \frac{\partial u_S}{\partial x} \bigg] .
\end{eqnarray}
These coupled differential equations were solved using Matlab routine {\tt pdepe}.
\\

\section{Results and discussion}
\label{sec:result}

\indent It was reported that the larger $\alpha$ or strong concentration gradient of small colloids ${\partial\phi_S}/{\partial z}$ drives the large colloids toward the bottom, causing the higher concentration of small colloids than large colloids near the freezing front. \cite{ZhouJiajia2017}
When the freezing front moves ahead, due to the adsorption of the interface between the solution and the ice, some colloidal particles would be trapped in the ice. 
The detail of the adsorption mechanism depends on many factors, such as the size of the colloids, functional groups on the surface, etc. 
To be general, here we consider two possibilities: (I) $k_S<k_L$ and (II) $k_S>k_L$. 
We study different values of $\alpha$ and ${\rm Pe}_S$, and numerically solve the coupled diffusion equations  (\ref{equ20}) and (\ref{equ21}). 
The total concentrations of the colloids in the ice are then calculated by integration.

\subsection{Case (I): $k_S < k_L$}
\label{sec:case1}

\begin{figure*}[htp]
	\centering
	\includegraphics[scale=0.8,draft=false]{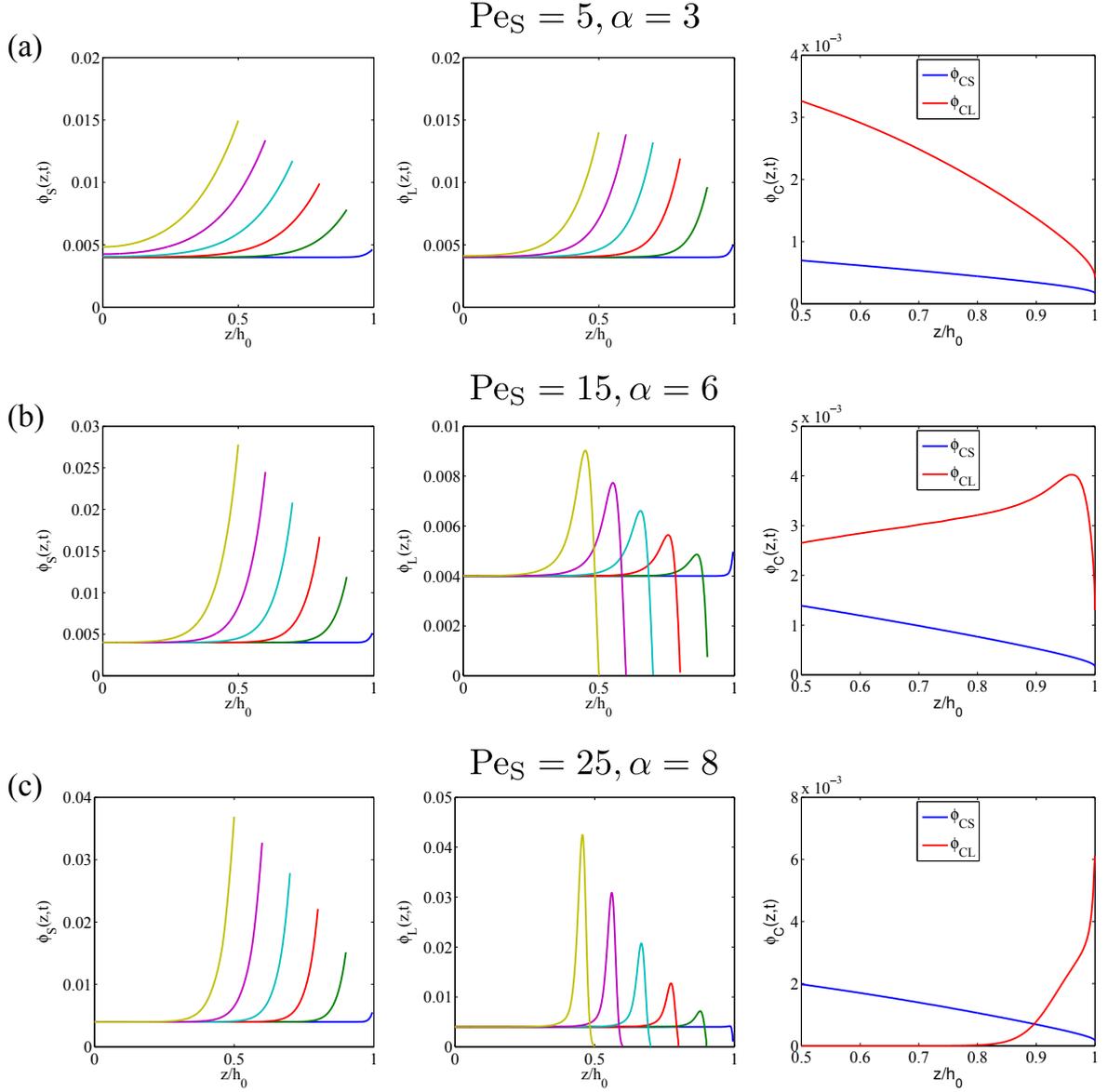}
	\caption{Volume fraction profiles of colloidal particles at different times $\tau$ = 0.005,0.1, 0.2, 0.3, 0.4, 0.5 (lines from bottom up) in the solution \change{for small colloids (the left column), for large colloids (the middle column), and in the ice (the right column; blue curves are for small colloids and red curves for large colloids)}. The initial concentrations are $\phi_{0S} =\phi_{0L} =0.004$. Parameters are $k_S = 0.04,k_L = 0.08$, satisfying the condition of $ k_S<k_L$, (a) ${\rm Pe}_S =5$ and $\alpha=3$, (b) ${\rm Pe}_S =15$ and $\alpha=6$,(c) ${\rm Pe}_S =25$ and $\alpha=8$.}
	\label{fig:2}
\end{figure*}

At first, we analyze the case $k_S(= 0.04) <k_L(= 0.08)$, i.e., the freezing front has a weaker adsorption effect on small colloids than large colloids. 
We select three representative sets of $\alpha$ and ${\rm Pe}_S$.
The concentration profiles are shown in Fig.~\ref{fig:2}. 
The left and middle columns show the concentration profiles in the solution for small colloids and large colloids, respectively.  
Each curve corresponds to a different time. 
The right column shows the concentration profile in the ice, and we stop the calculation when half of the initial solution is frozen. 
The initial concentrations of two type of colloids are small, $\phi_{0S}=\phi_{0L}=0.004$.

In Fig.~\ref{fig:2}(a), we show the case of ${\rm Pe}_S=5$ and $\alpha=3$. 
In this case,  the Peclet numbers for both types are great than unity, indicating that the diffusion is slow in comparison to the movement of the freezing front. 
Therefore, both types of colloids are accumulated near the ice front.
The concentration of the large colloids near the front is higher than that of small colloids because ${\rm Pe}_L = 3 {\rm Pe}_S > {\rm Pe}_S$. 
Since the large colloids also has a higher adsorption rate, $k_L > k_S$, more large colloids are trapped inside the ice phase, shown by the in-ice concentration profiles in the right panel of Fig.~\ref{fig:2}(a). 

When ${\rm Pe}_S=15$ and $\alpha=5$ [Fig.~\ref{fig:2}(b)], the small-on-top structure appears in the distribution of colloidal particles in the solution, similar to the case of evaporation-induced stratification.
In the ice phase, at the early times, most of colloids gather at the interface, and $k_S<k_L$ makes more large colloids to be stuck in the ice. 
At later times, the concentration gradient of small colloids becomes large, driving big colloids to the bottom eventually. 
The number of large colloids decreases quickly at the interface, so the concentration of large colloids in the ice will also be reduced. 
This leads to a broad peak in the concentration profile in the ice for large colloids. 

When the Peclet number and the size ratio increase further: ${\rm Pe}_S=25$ and $\alpha=8$ [Fig.~\ref{fig:2}(c)], the small-on-top structure fully develops in the solution due to large values of $\alpha$ and ${\rm Pe}_S$. 
In the ice phase, the concentration peak of large colloids becomes sharp and shifts to the right. 
At later times, the large colloids are depleted near the freezing front, resulting in that only small colloids are trapped inside the ice phase in the range of $0.5<z/h_0<0.8$.  

\vspace{0.2cm}
\begin{figure}[htp]
  \centering
  \includegraphics[width=0.8\columnwidth]{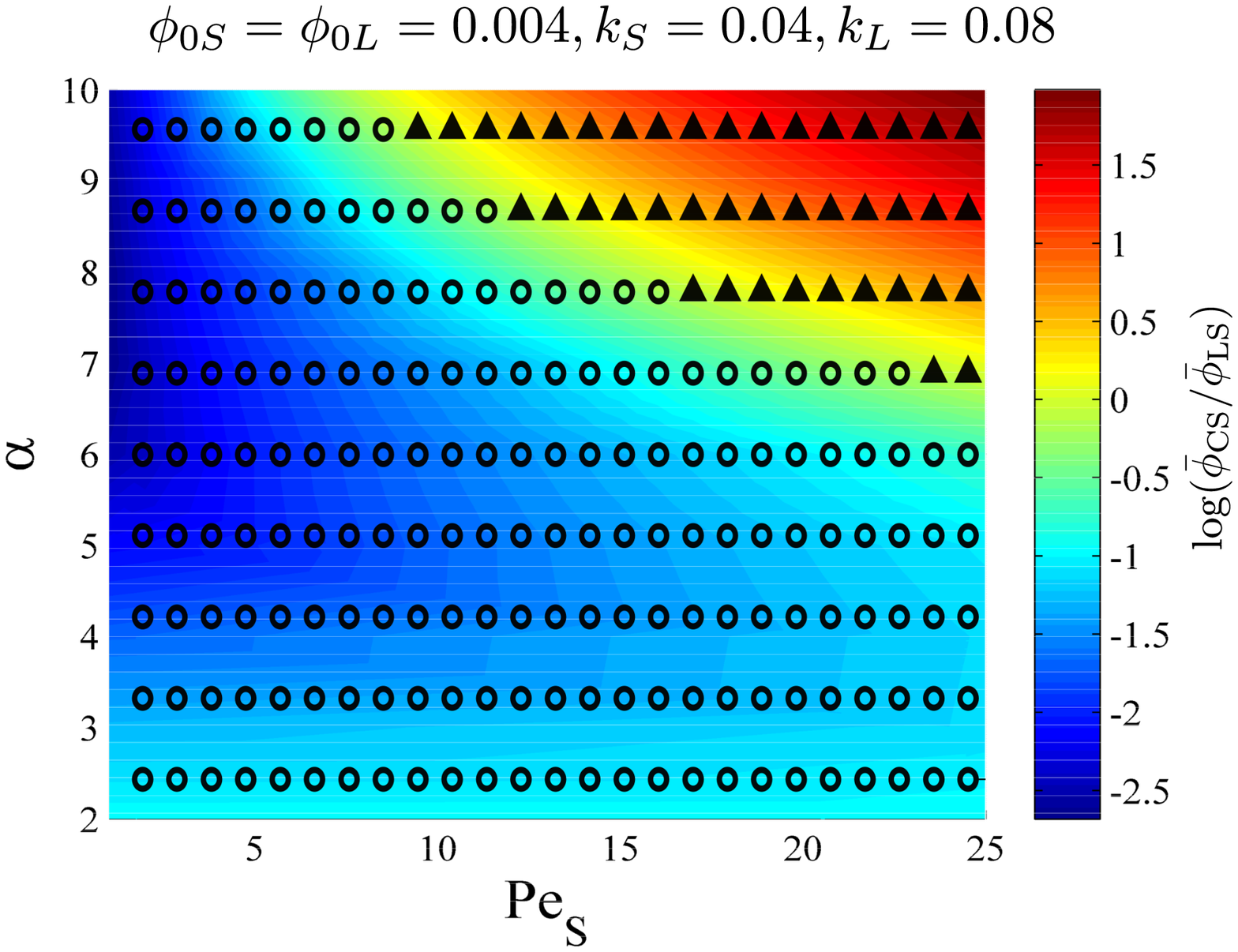}
  \caption{State diagram in the ${\rm Pe}_S$-$\alpha$ plane. The shade of color presents the concentration ratio of large colloids and small colloids at time $\tau$ = 0.5. Blue and circles show that the concentration of large colloids is higher than small colloids, while red and \change{filled triangles} show that large colloids is less than small colloids. The initial concentrations are $\phi_{0S} = \phi_{0L} =0.004$. Parameters of adsorption on colloidal particles are $k_S = 0.04, k_L = 0.08$, satisfying the condition of $k_S <k_L$.}
 	\label{fig:3}
 \end{figure}

In addition, the average concentration of two \change{types} of colloids in the ice can be obtained by integrating volume fraction profiles (the right column in Fig.~\ref{fig:2}) at time $\tau = 0.5$ 
\begin{equation}
  \change{\bar{\phi}_{Ci}} = \frac{2}{h_0} \int_{h_0/2}^{h_0} \phi_{Ci}(z) \, {\rm d}z .
\end{equation}  

This value is a good indicator for the efficiency of fractionation.   
Comparing the average concentration of small colloids with that of large colloids, a state diagram (Fig. \ref{fig:3}) in the ${\rm Pe}_S$-$\alpha$ plane is acquired.
The circles are for systems that the average concentration of large colloids is higher than small colloids, while squares show the case that large colloids is less than small colloids in the ice.
The color scale is the numerical value of $\log(\bar{\phi}_{SC} / \bar{\phi}_{LC})$.
The separation of the large/small colloids is more efficient in two parts of the state diagram: 
In the range when ${\rm Pe}_S$ is small while the $\alpha>5$, most particles in the ice are large colloids [Fig.~\ref{fig:2}(a)].
When the parameters ${\rm Pe}_S$ and $\alpha$ are both large, small colloids predominate in the ice phase. The later phenomenon corresponds to the Fig.~\ref{fig:2}(c).

\subsection{Case (II): $k_S>k_L$}
\label{sec:case2}

\begin{figure}[htp]
 	\centering
 	\includegraphics[scale=0.8,draft=false]{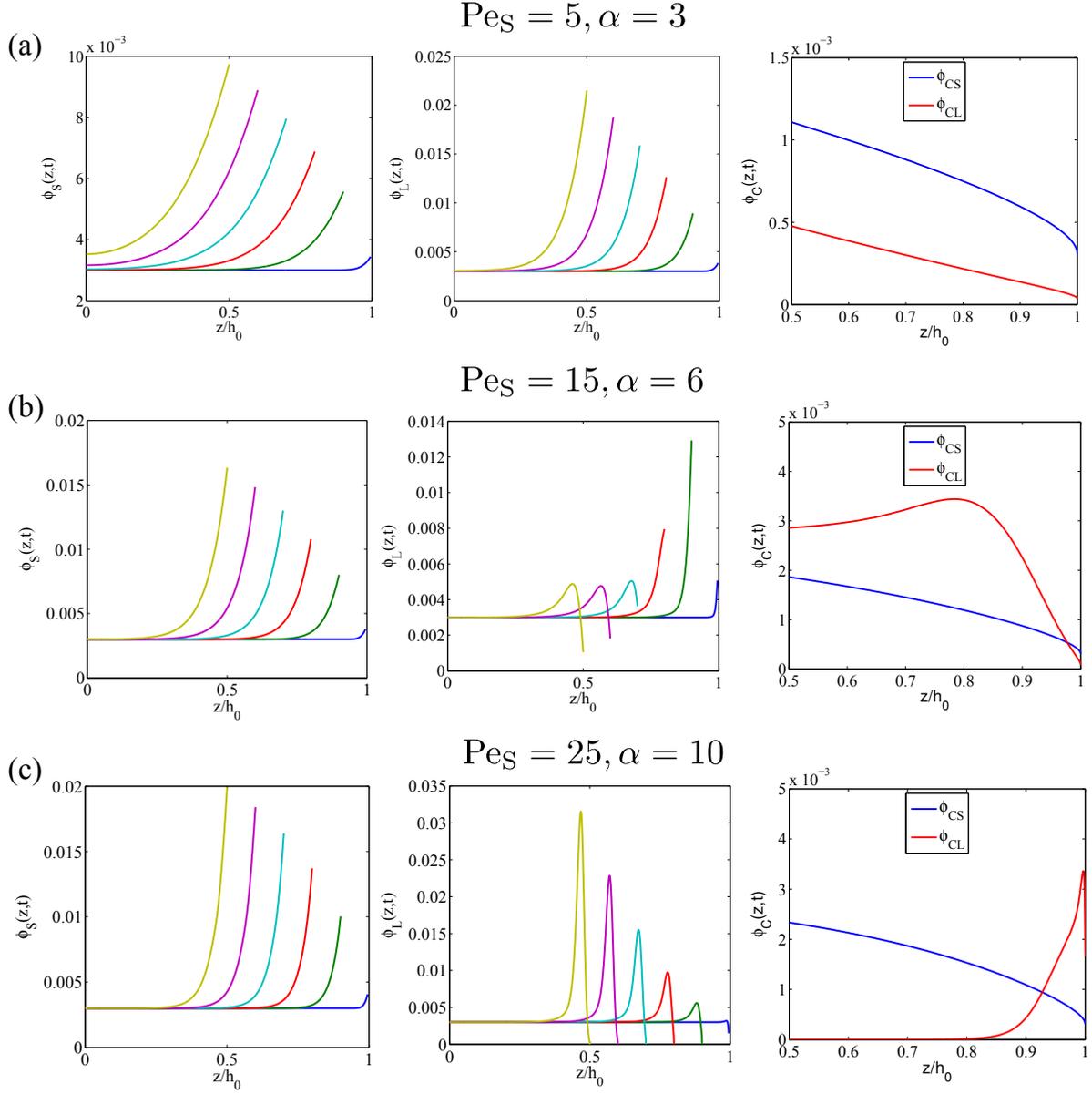}
 	\caption{Concentration fraction profiles of colloidal particles at different times $\tau$ =0.005, 0.1, 0.2, 0.3, 0.4, 0.5 (lines from bottom up) in the solution \change{for small colloids (the left column), for large colloids (the middle column), and in the ice (the right column; blue curves are for small colloids and red curves for large colloids)}. The initial concentrations are $\phi_{0S} = \phi_{0L} =0.003$. Parameters are $k_S =0.1, k_L = 0.01$, satisfying the condition of $k_S > k_L$, (a)${\rm Pe}_S =5$ and $\alpha=3$, (b) ${\rm Pe}_S =15$ and $\alpha=6$, (c) ${\rm Pe}_S =25$ and $\alpha=10$.}
 	\label{fig:4}
\end{figure}

We consider the other condition: $k_S(=0.1) > k_L(=0.01)$, i.e., the adsorption of small colloids at the freezing front is stronger than that of large colloids. We also select three typical examples via substitution of representative sets ($\alpha$, ${\rm Pe}_S$). 
Figure \ref{fig:4} shows concentration profiles of small colloids and large colloids at various times in the solution and in the ice. 
The initial concentrations of two type of colloids are $\phi_{0S}=\phi_{0L}=0.003$.

When ${\rm Pe}_S=5$ and $\alpha=3 $ [Fig.~\ref{fig:4}(a)], the concentration gradients for both colloids are small due to the relatively small values of the parameters. 
There is a small rise of the concentration near the freezing front on account of ${\rm Pe}_S$>1, meaning that the freezing front moves faster than the diffusion.
Since the adsorption rate for small colloids is larger than that of large colloids, the in-ice concentrations of small colloids is also larger than that of large colloids. 

A different scenario appears when ${\rm Pe}_S =15$ and $\alpha=6$ [Fig.~\ref{fig:4}(b)].
At the beginning, the large colloids gather at the interface because ${\rm Pe}_L > {\rm Pe}_S$.
In later times, the concentration gradient of the small colloids is large enough, pushing large colloids to the bottom. 
Similar to the intermediate state in Sect. \ref{sec:case1}, a broad peak in the concentration distribution of large colloids in the ice appears. 
In the ice phase, small colloids are trapped more than large colloids initially due to $k_S > k_L$. 
Later, due to the accumulation of larger colloids in the solution near the interface, the concentration of large colloids in the ice gradually increases. 
Finally, the concentration gradient of small colloids becomes stronger, pushing the large colloids away, and the large colloids trapped in the ice are reduced. 

When we increase the parameters even further, ${\rm Pe}_S =25$ and $\alpha=10$ [Fig.~\ref{fig:4}(c)], the small-on-top structure is formed at very early times.
Due to the depletion of the large colloids near the ice front, the peak of large colloids in the ice shifts to right and decreases sharply, and most of the colloids trapped are small ones.

\begin{figure}[htp]
  \centering
  \includegraphics[width=0.8\columnwidth]{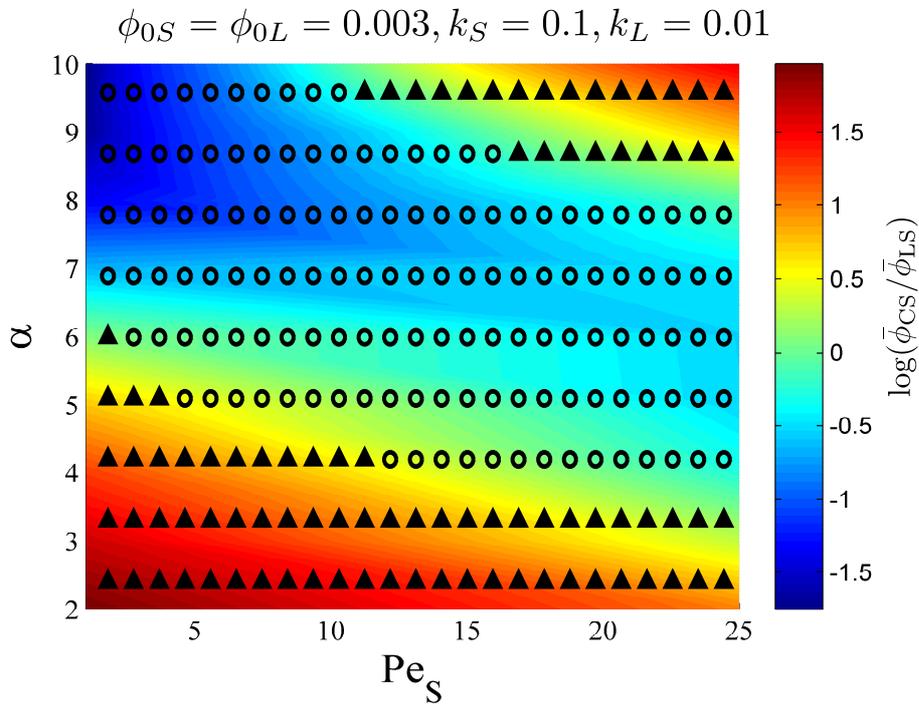}
  \caption{State diagram in the ${\rm Pe}_S$-$\alpha$ plane. The shade of color presents the concentration ratio of large colloids and small colloids at time $\tau$ = 0.5. Blue and circles show that the concentration of large colloids is higher than small colloids, while red and \change{filled triangles} show that large colloids is less than small colloids. The initial concentrations are $\phi_{0S}=\phi_{0L}=0.003$. Parameters are $k_S =0.1, k_L = 0.01$, satisfying the condition of $k_S > k_L$.}
  \label{fig:5}
\end{figure}

Similarly, we integrate volume fraction profiles in the ice at time $\tau=0.5$ to obtain the average concentration of two kind of colloids. 
A state diagram in the ${\rm Pe}_S$-$\alpha$ plane is acquired via comparing the average concentration of small colloids with that of large colloids. 
The results are shown in Fig.~\ref{fig:5}. 
In comparison to Fig.~\ref{fig:3}, there are two parts of the state diagram where the small colloids are trapped more in the ice than the large colloids (indicated by the red regions).
The cases in Fig.~\ref{fig:5} when ${\rm Pe}_S$ and $\alpha$ are both large is similar to that in Fig.~\ref{fig:3}; the small colloids are trapped predominately due to the small-on-top structure developed in the solution.
When both ${\rm Pe}_S$ and $\alpha$ are small, there are also more small colloids in the ice than the large colloids. 
However, the mechanism for this case is mainly that the adsorption rate of small colloids is large. 

\vspace{0.2cm}

We can compare our results with the experimental results \cite{GengHongya2017}.
The experimental data showed that with the increasing of the freezing speed (increase of ${\rm Pe}_S$), the GO sheets in the ice changes from mostly large sheets to predominately small sheets. 
We do not know exactly what the adsorption rates are, but if they are not too different, the state diagrams of Fig.~\ref{fig:3} and Fig.~\ref{fig:5} both show a similar trend at large $\alpha$ value. 
\\

\section{Summary}
\label{sec:summary}

In this article we develop a theoretical model to explain the fractionation of binary colloids at the ice growth front. 
By numerical solving the coupled diffusion equations, we demonstrate the mechanism of size fractionation. 
In the condition of $k_S  < k_L$, only when $\alpha$ and ${\rm Pe}_S$ are both large enough, there are more small colloids than the large colloids in the ice, while for other parameter values, there are more large colloids than small colloids. 
It attributes to the formation of small-on-top structure in the solution due to large $\alpha$ and ${\rm Pe}_S$ at early time, leading to that more small colloids are adsorbed in the ice. 
In the opposite condition of $k_S  > k_L$, when the values of $\alpha$ and ${\rm Pe}_S$ are both small or both large, more small colloids are trapped in the ice than large colloids, but at intermediate values of $\alpha$ and ${\rm Pe}_S$, the fractionation is opposite. 
The appearance of more small colloids trapped in ice at small values of $\alpha$ and ${\rm Pe}_S$ is due to the fact that larger adsorption rate for small colloids, while the small-on-top structure is not formed in the solution.


Our theoretical model demonstrates a delicate interplay between the adsorption rates and freezing speed. 
The size fractionation could be realized by careful control of the adsorption rates (by surface modification of GO sheets) and the directional freezing speed. 
When the case that more large colloids in the ice is desired, one should choose $k_S < k_L$ while keep the freezing slow. 
In the opposite case of more small colloids in the ice, one can resort to a large freezing speed, but not too fast to be in the regime of constitutional supercooling.

\change{There are a few limitations of our theoretical model and here we discuss possible extensions and improvements. 
We have employed a simple diffusion model and kept up to the second-order virial contribution to the free energy. 
These assumptions become questionable when the solution becomes concentrated. 
In general, more sophisticate free energy models can be used at high colloidal concentrations \cite{Carnahan1969, Hansen-Goos2006, Howard2017}.
For realistic GO nanosheets, the planar geometry is important and one has to consider the orientation degrees of freedom. 
There are also free energy models for disks and plates \cite{Zwanzig1954, Harnau2002, SuiJize2018}.
In practice, size fractionation needs to be performed for commercial GO solutions, which tend to be in the high concentration regime. 
In experiment, however, the range of the freezing rate for effective fractionation narrows at high concentrations \cite{GengHongya2017}, possibly due to the inhibition of the diffusion. 
Therefore, we have limited our studies to dilute solutions. 
More complicated separation methods may be preferred for concentrated solutions \cite{WangXiluan2011, ChenJi2015, ZhangWenjun2014, SunXiaoming2010, Lee2014}.
}

\change{
Another important effect is missing from our model is the hydrodynamic interaction.
We are using an implicit solvent model and the solvent is treated as a viscous background.
One can incorporate the hydrodynamic interactions by including the  concentration-dependence in the diffusion constant. 
There are many recent studies on the effect of hydrodynamics on the non-equilibrium colloidal dynamics, mostly by evaporation \cite{Sear2017, Sear2018, Statt2018, TangYanfei2018, TangYanfei2018a, TangYanfei2018b}.
One study on polymer solutions by Statt et al. \cite{Statt2018} showed that stratification happens in implicit solvent simulations, but not in explicit solvent simulations when hydrodynamic effects are included. 
One recent study on colloidal solutions by Tang et al. \cite{TangYanfei2018b}, however, observed the stratification for both explicit and implicit solvent simulations. 
Clearly there are more to be done to clarify the effect of hydrodynamic interactions.
}

\change{
In our model, the adsorption of the colloids to the ice phase is assumed to be in a quasi-equilibrium. 
This simplification thus allows us to use one simple parameter, $k_i$ in Eq. (\ref{equ14}), to characterize the adsorption rate. 
The real situation is more complicated. 
There are many forces involved when a solidification front approaches a colloidal particle:
(a) The disjoining force, originated from the thin layer of fluid in between the colloid's surface and the ice front. This force is the result of the temperature gradient and is of the van der Waals type \cite{Rempel1999, Rempel2001, YouJiaxue2018}. 
(b) The specific attraction or repulsion to the ice phase. In the case of GO, this can be adjusted by attaching functional groups on the GO surface. 
(c) The entropic force due to the concentration gradient of the colloids. 
(d) The hydrodynamic interaction between the colloids and the freezing front.
(e) The viscous friction force.
}

\change{
For large particles and fast freezing speed, the dynamics is governed by the balance of the disjoining force and the hydrodynamic force. 
This situation has been studied extensively \cite{Deville2009, Saint-Michel2017, YouJiaxue2018a, YouJiaxue2018b}, and the focus is on using the fast freezing to create complex composite materials \cite{Deville2006, ChengQunfeng2017}.
Here we purposely work in the regime of low freezing speed to avoid the instability of the freezing front. 
In this case, especially for small colloids, the dynamics is determined by the balance of the specific interaction, entropic force, and the viscous friction.
The specific interaction can be the results of hydrophobic interaction, the hydrogen-bonding interaction, etc., mostly of the short-range type. 
A recent experimental work on the directional freezing of emulsions demonstrated that long-range interactions are also important \cite{Dedovets2018}.
In our model, the effect of these interactions are grouped into one phenomenological parameter $k_i$. 
More works have to be done to determine the connection between the $k_i$ parameter and the detailed interaction between the colloids/nanosheets and the freezing front.
}



\begin{acknowledgments}
This work was supported by the National Natural Science Foundation of China (21504004, 21574006, 21622401, 21774004) and the 111 Project (B14009). 
M.D. acknowledges the financial support of the Chinese Central Government in the Thousand Talents Program.
\end{acknowledgments}

\bibliography{freezing}

\end{document}